\documentclass[notitlepage]{article}
\usepackage[T1]{fontenc}
\usepackage[american]{babel}
\usepackage[dvips]{graphicx}
\usepackage{color}
\usepackage{vmargin}

\begin{document}

\begin{center}
\textbf{
COVARIANT FORMULATION\ OF\ ELECTROMAGNETIC\ 4-MOMENTUM\ IN\ TERMS\ OF\
4-VECTORS $E^{\alpha }$ AND $B^{\alpha }$
}%
\bigskip

Tomislav Ivezi\'{c}\bigskip

Ru\textit{{%
\mbox
 {\it{d}\hspace{-.15em}\rule[1.25ex]{.2em}{.04ex}\hspace{-.05em}}}er Bo\v
{s}kovi\'{c} Institute, P.O.B. 180, 10002 Zagreb, Croatia}

ivezic@rudjer.irb.hr\bigskip
\end{center}


\begin{abstract}
The fundamental difference between the true transformations (TT)
and the apparent transformations (AT) is explained. The TT refer to the same
quantity, while the AT refer, e.g., to the same measurement in different
inertial frames of reference. It is shown that the usual transformations of
the three-vectors ${\bf E}$ and ${\bf B}$ are - the AT. The covariant
electrodynamics with the four-vectors $E^{\alpha }$ and $B^{\alpha }$ of the
electric and magnetic field is constructed. It is also shown that the
conventional synchronous definitions of the electromagnetic energy and
momentum contain both, the AT of the volume, i.e., the Lorentz contraction,
and the AT of ${\bf E}$ and ${\bf B}$, while Rohrlich's expressions contain
only the AT of ${\bf E}$ and ${\bf B}$. A manifestly covariant expression
for the energy-momentum density tensor and the electromagnetic 4-momentum is
constructed using $E^{\alpha }$ and $B^{\alpha }$. The ''4/3'' problem is
discussed and it is shown that all previous treatments either contain the AT
of the volume, or the AT of ${\bf E}$ and ${\bf B}$, or both of them. In our
approach all quantities are four-dimensional spacetime tensors whose
transformations are the TT.
\end{abstract}

\section{INTRODUCTION}

The correct expressions for the electromagnetic energy and momentum and
their transformation properties are topics of repeated discussions in this
journal and elsewhere. Two approaches to these problems can be
distinguished. The main points of both approaches are fairly well exposed in
the Rohrlich- Boyer discussion \cite{RochD,Boyer} and in the subsequent
Comment \cite{camp}.

In 1966. Rohrlich \cite{rochnc} introduced the notions of the true and
apparent transformations of physical quantities. The transformations of the
four-dimensional (4D) spacetime tensors referring to the same quantity
considered in different inertial frames of reference (IFRs) are the true
transformations (TT). They are in full agreement with the special relativity
as the theory of 4D spacetime with pseudo-Euclidean geometry. An example of
the TT are the Lorentz transformations (LT) of 4D tensor quantities. On the
contrary the transformations which do not refer to the same physical
quantity, e.g., the transformations which refer to the same measurement in
different IFRs, are the apparent transformations (AT). Rohrlich \cite{rochnc}
and Gamba \cite{gamba} discussed different examples of the TT and the AT and
explicitly showed that the Lorentz contraction of length (volume) belongs to
the class of - the AT.

The fundamental difference between the AT and the TT of physical quantities
is previously mainly overlooked. The importance of that difference is
emphasized in this paper and it is shown here that the usual transformations
of the electric ${\bf E}$ and the magnetic ${\bf B}$ fields as the
three-vectors (3-vectors) also belong to the class of - the AT. The
4-vectors $E^{\alpha }$ and $B^{\alpha }$ are introduced instead of ${\bf E}$
and ${\bf B}$ and the covariant Maxwell equations are formulated in terms of 
$E^{\alpha }$ and $B^{\alpha }$. This alternative covariant formulation with 
$E^{\alpha }$ and $B^{\alpha }$ is equivalent to the usual covariant
electrodynamics with the electromagnetic field tensor $F^{\alpha \beta }$,
(see also \cite{ivez12}).

It has to be pointed out that our covariant approach with $E^{\alpha }$ and $%
B^{\alpha }$ does not make use of the intermediate electromagnetic
4-potential $A^{\mu }$, and thus dispenses with the need for gauge
conditions.

The existence of the difference between the AT and the TT causes that one
can speak about two forms of relativity: the ''AT relativity'' and the ''TT
relativity.'' The former is the conventional special relativity based on
Einstein's relativity of simultaneity and on the synchronous definition of
length, i.e., on the AT of length and time, and, as shown here, on the AT of
the electric and magnetic 3-vectors ${\bf E}$ and ${\bf B}$. The ''TT
relativity,'' or, equivalently, the covariant formulation of relativity, is
based on the TT of physical quantities as 4D spacetime tensors, i.e., on the
covariant definition of length, and the covariant electrodynamics with $%
F^{\alpha \beta }$ or, equivalently, as shown here, with 4-vectors $%
E^{\alpha }$ and $B^{\alpha }$.

The alternative covariant approach is used to discuss different definitions
of the electromagnetic momentum and energy, and also to discuss the related
''4/3'' problem in the electromagnetic mass of the electron. First in Sec. 2
the AT and the TT of length (volume) and of the electric and magnetic fields
are considered. It is also shown that, contrary to von Laue's theorem, an
integral of a symmetric tensor of second rank over the hyperplane $t=const.$
does not form a true 4-vector, since the transformation of that integral
from an IFR $S$ to another IFR $S^{\prime }$ is - an AT. These results are
used in all other Sections. In Sec. 3 it is proved that the conventional
definitions of the electromagnetic energy and momentum are not covariant
definitions, since they are synchronous definitions, which use the AT of
volume elements, i.e., the Lorentz contraction, and the AT for ${\bf E}$ and 
${\bf B}$. Rohrlich's approach is discussed in Sec. 4. It is shown that
neither Rohrlich's relations for the electromagnetic energy and momentum
define the true 4-vector, since the electromagnetic field tensor $F^{\alpha
\beta }$ is expressed in terms of ${\bf E}$ and ${\bf B}$ and the AT for $%
{\bf E}$ and ${\bf B}$ are used. In Sec. 5 we replace ${\bf E}$ and ${\bf B}$
with $E^{\alpha }$ and $B^{\alpha }$ and compare the obtained expressions
with Rohrlich's relations containing ${\bf E}$ and ${\bf B.}$ Furthermore,
the 4/3 problem in the electromagnetic mass is discussed in this Section. In
Sec. 6 some recent treatments of the electromagnetic energy and momentum and
of the 4/3 problem are discussed and it is shown that in all of them either
the AT of volume elements, or the AT of ${\bf E}$ and ${\bf B}$, or both of
them, are used. This means that none of these treatments do conform with the
''TT relativity.''

\section{TRUE AND APPARENT TRANSFORMATIONS}

According to the ''modern'' point of view the special relativity is the
theory of 4D spacetime with pseudo-Euclidean geometry. Quantities of
physical interest, both local and nonlocal, are represented in the special
relativity by spacetime tensors, i.e., as covariant quantities. {\em The
laws of physics are written in the special relativity in a manifestly
covariant way as tensorial equations. The geometry of the spacetime is
defined by the invariant infinitesimal spacetime distance} $ds$\ {\em of two
neighboring points}, $ds^{2}=g_{\mu \nu }dx^{\mu }dx^{\nu }$($g_{\mu \nu }$\
is the metric tensor; Greek indices run from $0$\ to $3$, latin indices run
from $1$\ to $3$, and repeated indices imply summation).{\em \ The laws of
physics written as tensorial equations with 4D spacetime tensors in an IFR
will have the same form in some other IFR, i.e., in new coordinates, if new
and old coordinates are connected by those coordinate transformations that
leave the interval} $ds$, {\em and thus the pseudo-Euclidean geometry of the
spacetime, unchanged.} This means that in the reference frames that are
connected by such coordinate transformations all physical phenomena will
proceed in the same way, (taking into account the corresponding initial and
boundary conditions), and thus there is no physical difference between them,
(the principle of relativity). The transformations that leave $ds$ unchanged
also transform a physical quantity represented by 4D spacetime tensor from
an IFR $S$ to another IFR $S^{\prime }$; the same quantity is considered in $%
S$ and $S^{\prime }$. Such transformations are the TT and an example of them
are, as already said, the LT between IFRs. The meaning of the same quantity
Rohrlich \cite{rochnc} expresses in the following way: ''A quantity is
therefore physically meaningful (in the sense that it is of the same nature
to all observers) if it has tensorial properties under Lorentz
transformations.'' Similarly Gamba \cite{gamba}, when discussing the
sameness of a physical quantity (for example, the quantity $A_{\mu
}(x_{\lambda },X_{\lambda })$, which is a function of two points $x_{\lambda
}$ and $X_{\lambda }$) for different IFRs $S$ and $S^{\prime }$, declares:
''The quantity $A_{\mu }(x_{\lambda },X_{\lambda })$ for $S$ is the same as
the quantity $A_{\mu }^{\prime }(x_{\lambda }^{\prime },X_{\lambda }^{\prime
})$ for $S^{\prime }$ when all the primed quantities are obtained from the
corresponding unprimed quantities through Lorentz transformations (tensor
calculus).''

It has to be noted that, in principle, one can choose any reference frame in
which some physical quantity is covariantly defined, and then the same
quantity is considered in all other reference frames, transforming all parts
of that quantity by the TT. The whole physics will not depend on the chosen
frame. However, the most convenient choice for systems with rest mass is the
rest frame, since in that frame one retains the similarity with the
prerelativistic physics.

\subsection{True and Apparent Transformations of Length (Volume)}

The definition of length (volume) which is in accordance with all the above
consideration is - the covariant definition of length (see, e.g., Ref. 7 and
references therein, or for the volume \cite{rochnc,gamba}). The invariant
spacetime length (the Lorentz scalar) is formed as $l=(l^{\mu }l_{\mu
})^{1/2}$ where $l_{\mu }$ is the distance 4-vector between two spatial
points A and B on the (moving) object, $l^{\mu }=x_{B}^{\mu }-x_{A}^{\mu },$ 
$x_{A,B}^{\mu }$ are the position 4-vectors in some IFR $S$. (We use
Cartesian space coordinates $x^{i}$ and time $t(x^{0}\equiv ct)$. Without
loss of generality we work in an IFR with the Minkowski metric tensor $%
g_{\mu \nu }$=$diag(-1,1,1,1)$.) The observers in all other IFRs will look
at the same events but associating with them different coordinates; it is
the essence of the covariant description. The LT of the position 4-vectors,
of the distance 4-vector, and of the invariant spacetime length are all the
TT.

In contrast to the covariant definition of length and the TT of the
spacetime tensors {\em the synchronous definition of length}, introduced by
Einstein \cite{einst}, {\em defines length as the spatial distance between
two points on the (moving) object measured by simultaneity in the rest frame
of the observer}. Let $l_{0}$ be the rest length, or the three-dimensional
volume $dV_{0}$, measured by simultaneity in $S_{0}$, the rest frame of the
object, at some $t_{0}=a$. Then the length $l^{\prime }$ (the 3D volume $%
dV^{\prime }$) determined simultaneously in some IFR $S^{\prime }$, at some $%
t^{\prime }=b$, is Lorentz contracted, $l^{\prime }=l_{0}(1-\beta
^{2})^{1/2} $, ($dV^{\prime }=dV_{00}(1-\beta ^{2})^{1/2}$), ($\beta =V/c$, $%
V$ is the relative velocity of $S_{0}$ and $S^{\prime }$. It has to be
emphasized that $t_{0}=a$ in $S_{0}$ and $t^{\prime }=b$ in $S^{\prime }$
are not related by the LT or any other coordinate transformation. The time
component is not transformed in the Lorentz contraction. This means that the
observers in $S_{0}$ and $S^{\prime }$ do not look at the same events. Hence
the quantities $l_{0}$ and $l^{\prime }$ synchronously determined in the
rest frame of the observer{\em \ refer to the same measurement in }$S_{0}$%
{\em \ and }$S^{\prime }$ and not to the same 4D tensor quantity. {\em %
Therefore the Lorentz contraction as the transformation connecting them is -
the AT.}

\subsection{Von Laue's theorem}

The existence of the fundamental difference between the TT and the AT
enables us to examine an important theorem (it is sometimes called ''von
Laue's theorem''), which is widely used in the theory of relativity and in
the quantum theory of fields. This theorem roughly states: The necessary and
sufficient condition for the hyperplane integral of a symmetric tensor of
second rank $T$ ($\int_{\Sigma }T^{\mu \nu }(x)d^{3}\sigma _{\nu }(x)$) to
be independent of the orientation of that hyperplane ($\Sigma $) is that $T$
be divergence-free, $\partial _{\nu }T^{\mu \nu }=0$, (local conservation
law); and if that integral is orientation-independent, then it can be
written as $\int_{t=a}T^{\mu 0}({\bf r},t)d^{3}x$ and it is 1) independent
of time, ($\int_{t=a}T^{\mu 0}({\bf r},t)d^{3}x=\int_{t=b}T^{\mu 0}({\bf r}%
,t)d^{3}x$), and 2) it is a 4-vector, (the proof of that theorem can be
found in, e.g., Ref .9, Sec. 5.8, and Ref. 10). We are not interesting in
the time independence of that integral but in its 4-vector character. In the
proof of 2) the integrals are taken over the hypersurfaces $t=a$ in $S$ and $%
t^{\prime }=b$ in $S^{\prime }$. According to the previous discussion about
the sameness of a physical quantity for different IFRs, and the related
consideration about the TT and the AT, we conclude: {\em The integral }$%
\int_{t=a}T^{\mu 0}({\bf r},t)d^{3}x${\em \ cannot be a true 4-vector from
the ''TT viewpoint.''} Namely, {\em this integral does not refer to the same
quantity considered from different IFRs, (which are connected by the LT),
since the hyperplanes taken at }$t=a${\em \ in }$S${\em \ and }$t^{\prime
}=b ${\em \ in }$S^{\prime }${\em \ are not related in any way.} Thus, e.g.,
the primed elements of volume $d^{3}x^{\prime }$ in the integral in $%
S^{\prime }$ ($\int_{t^{\prime }=b}T^{^{\prime }\mu 0}({\bf r}^{\prime
},t^{\prime })d^{3}x^{\prime }$) are not obtained by the LT from $S$, than
they are simply the elements of volume of {\em an arbitrary chosen
hypersurface }$t^{\prime }=b${\em \ in }$S^{\prime }${\em . The
transformation connecting the integrals in }$S${\em \ and }$S^{\prime }${\em %
\ is - an AT, since not all parts of that compound physical quantity are
transformed by the LT from }$S${\em \ to }$S^{\prime }$. The situation with
the transformation of that integral is quite similar to the already
discussed Lorentz contraction as an AT. We see that, contrary to von Laue's
theorem, the vanishing of the divergence of $T$ does not assure that the
mentioned integral is a true 4-vector. The considerations and the proofs
given in the previous literature neglected that the ''TT relativity''
demands the same physical quantity to be considered from different IFRs,
i.e., that all parts of that quantity have to be transformed by the LT from $%
S$ to $S^{\prime }$.

\subsection{The Proof that the Transformations of ${\bf E}$ and ${\bf B}$
are - the AT}

As a next example we consider the transformations of ${\bf E}$ and ${\bf B}$%
. It is generally believed that the covariant formulation of the
electrodynamics with $F^{\alpha \beta }$ and the usual formulation with $%
{\bf E}$ and ${\bf B}$ are equivalent, and therefore that the usual
transformations of ${\bf E}$ and ${\bf B}$ are actually the TT. However we
show that these transformations also belong to the class of - the AT, (see
also\cite{ivez12}). In the modern derivation of the transformation relations
for ${\bf E}$ and ${\bf B}$, (see, e.g., Ref. 11 Sec. 11.10, Ref. 9 Sec.
3.3.), one identifies, in some IFR $S$, the components of the 3-vectors $%
E_{i}$ and $B_{i}$ with the components of $F^{\alpha \beta }$ as $%
E_{i}=F^{0i},$ and $B_{i}=(1/c)^{*}F^{0i}$ in order to get in that IFR the
usual Maxwell equations, 
\begin{eqnarray*}
\nabla {\bf E}({\bf r},t) &=&\rho ({\bf r},t)/\varepsilon _{0},\quad \nabla
\times {\bf E}({\bf r},t)=-\partial {\bf B}({\bf r},t)/\partial t, \\
\nabla {\bf B}({\bf r},t) &=&0,\quad \nabla \times {\bf B}({\bf r},t)=\mu
_{0}{\bf j}({\bf r},t)+\mu _{0}\varepsilon _{0}\partial {\bf E}({\bf r}%
,t)/\partial t,
\end{eqnarray*}
from the covariant Maxwell equations with $F^{\alpha \beta }$ and its dual $%
^{*}F^{\alpha \beta }$ 
\[
\partial _{\alpha }F^{a\beta }=-j^{\beta }/\varepsilon _{0}c,\quad \partial
_{\alpha }\ ^{*}F^{\alpha \beta }=0 
\]
where $^{*}F^{\alpha \beta }=-(1/2)\varepsilon ^{\alpha \beta \gamma \delta
}F_{\gamma \delta }$ and $\varepsilon ^{\alpha \beta \gamma \delta }$ is the
totally skew-symmetric Levi-Civita pseudotensor. (Note that in the covariant
formulation $F^{\alpha \beta }$ is the primary quantity; it is the solution
of the covariant Maxwell equations, or the corresponding wave equation $%
\partial ^{\sigma }\partial _{\sigma }F_{\alpha \beta }-(1/\varepsilon
_{0}c)(\partial _{\beta }j_{\alpha }-\partial _{\alpha }j_{\beta })=0,$ and
it conveys all the information about the electromagnetic field. $F^{\alpha
\beta }$ is generally given as 
\[
F^{\alpha \beta }(x^{\mu })=(2k/i\pi c)\int \left\{ \frac{\left[ j^{\alpha
}(x^{\prime \mu })(x-x^{\prime })^{\beta }-j^{\beta }(x^{\prime \mu
})(x-x^{\prime })^{\alpha }\right] }{\left[ (x-x^{\prime })^{\sigma
}(x-x^{\prime })_{\sigma }\right] ^{2}}\right\} d^{4}x^{\prime }, 
\]
where $x^{\alpha },x^{\prime \alpha }$ are the position 4-vectors of the
field point and the source point respectively, and $k=1/4\pi \varepsilon
_{0}.)$ After transforming by the LT the covariant Maxwell equations to the $%
S^{\prime }$ frame one finds $\partial _{\alpha }^{\prime }F^{\prime a\beta
}=-j^{\prime \beta }/\varepsilon _{0}c,\quad \partial _{\alpha }^{\prime }\
^{*}F^{\prime \alpha \beta }=0.$ The transformations of all quantities in
the covariant Maxwell equations are - the TT. The covariant Maxwell
equations do not change their form on the LT embodying in that way the
principle of relativity. Then one again identifies the 3-vectors $%
E_{i}^{\prime }$ and $B_{i}^{\prime }$ with the components of $F^{\prime
\alpha \beta }$ {\em in the same way as in} $S$, i.e., $E_{i}^{\prime
}=F^{\prime 0i},$ and $B_{i}^{\prime }=(1/c)^{*}F^{\prime 0i}$ in order to
obtain the usual Maxwell equations (in the three-vector form) from the
transformed covariant Maxwell equations. This procedure then gives the
connection between the 3-vectors $E_{i}^{\prime },$ $B_{i}^{\prime }$ in $%
S^{\prime }$ and $E_{i},$ $B_{i}$ in $S$ as 
\begin{eqnarray}
E_{i} &=&\Gamma (E_{i}^{\prime }-c\varepsilon _{ijk}\beta _{j}B_{k}^{\prime
})-((\Gamma -1)/\beta ^{2})\beta _{i}(\beta _{k}E_{k}^{\prime })  \nonumber
\\
B_{i} &=&\Gamma (B_{i}^{\prime }-(1/c)\varepsilon _{ijk}\beta
_{j}E_{k}^{\prime })-((\Gamma -1)/\beta ^{2})\beta _{i}(\beta
_{k}B_{k}^{\prime }),  \label{eitran}
\end{eqnarray}
where $\beta =V/c$ and $\Gamma =(1-\beta ^{2})^{-1/2}.$ (The components of
the 3-vector fields ${\bf E}$ and ${\bf B}$, and of the 3-velocity ${\bf V}$
are written with lowered (generic) subscripts, since they are not the
spatial components of the 4-vectors. This refers to the third-rank
antisymmetric $\varepsilon $ tensor too. The super- and subscripts are used
only on the components of the 4-vectors or tensors.) According to (\ref
{eitran}) the components of, e.g., the 3-vector ${\bf E}^{\prime }$ in $%
S^{\prime }$ are determined by the components of both 3-vectors ${\bf E}$
and ${\bf B}$ in $S$. Obviously ${\bf E}$ in $S$, {\em measured by the
observers at rest in }$S$, and ${\bf E}^{\prime }$ in $S^{\prime }$, {\em %
measured by the observers at rest in} $S^{\prime }$, (the same holds for $%
{\bf B}$ and ${\bf B}^{\prime }$), {\em do not refer to the same quantity
considered in }$S${\em \ and }$S^{\prime }${\em , but to the same measurement%
}, as in the Lorentz contraction, and consequently the 3-vectors ${\bf E}$
and ${\bf E}^{\prime }$ (${\bf B}$ and ${\bf B}^{\prime }$) are not
connected by the TT than by the AT, Eq. (\ref{eitran}).

\subsection{The 4-vectors $E^{\alpha }$ and $B^{\alpha }$ and the AT of $%
{\bf E}$ and ${\bf B}$}

Although $F^{\alpha \beta }$ contains all the information about the
electromagnetic field one can introduce the 4-vectors $E^{\alpha }$ and $%
B^{\alpha }$ instead of the usual 3-vectors ${\bf E}$ and ${\bf B}$. The $%
E^{\alpha }$ and $B^{\alpha }$ are well defined quantities from the ''TT
viewpoint'' and they are defined by means of $F^{\alpha \beta }$ as 
\begin{equation}
E^{\alpha }=(1/c)F^{\alpha \beta }v_{\beta ,}\quad B^{\alpha
}=(1/c^{2})^{*}F^{\alpha \beta }v_{\beta }.  \label{veef}
\end{equation}
The $E^{\alpha }$ and $B^{\alpha }$ are the electric and magnetic field
4-vectors measured by an observer moving with 4-velocity $v^{\alpha }$ in an
arbitrary reference frame.

We note that in this paper $E^{\alpha }$ and $B^{\alpha }$ are defined in
the same way as in \cite{wald}, i.e., taking that $v^{\alpha }$ is the
4-velocity of a family of observers who measures the fields. But it is
considered in \cite{wald} that $E^{\alpha }$ and $B^{\alpha }$ are necessary
only for noninertial and curved spacetimes and not for IFRs. However the
fundamental result that the usual transformations of ${\bf E}$ and ${\bf B}$%
, Eq.(\ref{eitran}), belong to the class of the AT necessitates the
introduction of the 4-vectors $E^{\alpha }$ and $B^{\alpha }$ even for IFRs.
The definition (\ref{veef})can be compared with the definitions of $%
E^{\alpha }$ and $B^{\alpha }$ in, e.g., Ref. 13, where the covariant
electrodynamics in the moving medium is considered and $v^{\alpha }$ is the
4-velocity of the medium, or in Ref. 14, where the physical meaning of $%
v^{\alpha }$ is unspecified - it is any unitary 4-vector. The reason for
such choice of $v^{\alpha }$ in \cite{espos} is that there $E^{\alpha }$ and 
$B^{\alpha }$ are introduced as the ''auxiliary fields,'' while ${\bf E}$
and ${\bf B}$ are considered as the physical fields. In our alternative
covariant approach the situation is just the opposite; $E^{\alpha }$ and $%
B^{\alpha }$ are the real physical fields, which are correctly defined and
measured in 4D spacetime, while the 3-vectors ${\bf E}$ and ${\bf B}$ are
not correctly defined in 4D spacetime from the ''TT viewpoint.''

The introduction of $E^{\alpha }$ and $B^{\alpha }$ enables us to better
explain why the transformations of ${\bf E}$ and ${\bf B}$, (\ref{eitran}),
are the AT. Namely, {\em in the mentioned modern derivation of (\ref{eitran}%
) two different families of observers who measure the electric field are
considered, one at rest in the IFR }$S${\em , for which }$v^{\alpha }=(c,%
{\bf 0})${\em , and another one at rest in the IFR S', for which again }$%
v^{\prime \alpha }=(c,{\bf 0})${\em , and these observers in }$S${\em \ and }%
$S^{\prime }${\em \ are not related in any way.} Such assumptions for $%
v^{\alpha }$ and $v^{\prime \alpha }$ mean that {\em one does not consider
the same physical quantity in }$S${\em \ and }$S^{\prime }${\em ,} but that
two different quantities $E^{\alpha }$ and $\Xi ^{\prime \alpha }$ (in which 
$v^{\alpha }$ is not transformed) are considered in $S$ and $S^{\prime }$,
respectively. The quantity $\Xi ^{\prime \alpha }$ has nothing in common
with the electric field $E^{\alpha }$. This means that the transformations (%
\ref{eitran}) are the AT; {\em they refer to the same measurement in two
IFRs and not to the same physical quantity as required by the ''TT
relativity.''} The TT referring to the same physical quantity are the LT of $%
E^{\alpha }$ and $B^{\alpha }$.

It has to be mentioned here that although Rohrlich, \cite{rochnc,RochD} and
Gamba \cite{gamba} correctly insist on covariant definitions of various
physical quantities, they also did not notice that equations (\ref{eitran})
are the AT that do not refer to the same physical quantity in two IFRs.

\subsection{Alternative Covariant Formulation with $E^{\alpha }$ and $%
B^{\alpha }$}

Using $E^{\alpha }$ and $B^{\alpha }$ one can construct the covariant
formulation of electrodynamics, which is equivalent to the usual covariant
formulation with $F^{\alpha \beta }$. For that one needs the inverse
relations to the relations (\ref{veef}) in which $F^{\alpha \beta }$ will be
expressed by means of $E^{\alpha }$ and $B^{\alpha }$, and $v^{\alpha }$,
(compare with Refs. 13 and 14 taking into account the above mentioned
remarks about the meaning of $E^{\alpha },$ $B^{\alpha }$, and $v^{\alpha }$
in these works). The inverse relations are 
\begin{eqnarray}
F^{\alpha \beta } &=&(1/c)\delta _{\mu \nu }^{\alpha \beta }v^{\mu }E^{\nu
}+\varepsilon ^{\alpha \beta \mu \nu }B_{\mu }v_{\nu },  \nonumber \\
^{\ast }F^{\alpha \beta } &=&\delta _{\mu \nu }^{\alpha \beta }v^{\mu
}B^{\nu }+(1/c)\varepsilon ^{\alpha \beta \mu \nu }v_{\mu }E_{\nu },
\label{vzF}
\end{eqnarray}
where $\delta _{\mu \nu }^{\alpha \beta }=\delta _{\mu }^{\alpha }\delta
_{\nu }^{\beta }-\delta _{\nu }^{\alpha }\delta _{\mu }^{\beta }.$ The
4-vectors $E^{\alpha }$ and $B^{\alpha }$ satisfy the conditions $v_{\alpha
}E^{\alpha }=v_{\beta }B^{\beta }=0$, as can be checked from (\ref{veef})
and (3). Substituting (3) into the covariant Maxwell equations with $%
F^{\alpha \beta }$ we obtain the covariant Maxwell equations with $E^{\alpha
}$ and $B^{\alpha }$, 
\begin{eqnarray*}
\partial _{\alpha }(\delta _{\mu \nu }^{\alpha \beta }v^{\mu }E^{\nu
})+c\partial _{\alpha }(\varepsilon ^{\alpha \beta \mu \nu }B_{\mu }v_{\nu
}) &=&-j^{\beta }/\varepsilon _{0}, \\
\partial _{\alpha }(\delta _{\mu \nu }^{\alpha \beta }v^{\mu }B^{\nu
})+(1/c)\partial _{\alpha }(\varepsilon ^{\alpha \beta \mu \nu }v_{\mu
}E_{\nu }) &=&0.
\end{eqnarray*}
The relations (3) transform the covariant formulation with $F^{\alpha \beta
} $ and $^{*}F^{\alpha \beta }$ to the covariant formulation with $E^{\alpha
}$ and $B^{\alpha }$, while the relations (\ref{veef}) do the reverse
transformations, (see also \cite{ivez12}).

If one takes that in an IFR $S$ the observers who measure $E^{\alpha }$ and $%
B^{\alpha }$ are at rest, i.e., $v^{\alpha }=(c,{\bf 0})$, then $%
E^{0}=B^{0}=0$, and one can derive from the covariant Maxwell equations with 
$E^{\alpha }$ and $B^{\alpha }$ the Maxwell equations which contain only the
space parts $E^{i}$ and $B^{i}$ of $E^{\alpha }$ and $B^{\alpha }$, e.g.,
from the first covariant Maxwell equation one easily finds $\partial
_{i}E^{i}=j^{0}/\varepsilon _{0}c$. We see that the Maxwell equations
obtained in such a way are of the same form as the usual Maxwell equations
with ${\bf E}$ and ${\bf B}$. From the above consideration one concludes
that all the results obtained in a given IFR $S$ from the usual Maxwell
equations with ${\bf E}$ and ${\bf B}$ remain valid in the covariant
formulation with the 4-vectors $E^{\alpha }$ and $B^{\alpha }$ but only for
the observers who measure the fields $E^{\alpha }$ and $B^{\alpha }$ and are
at rest in the considered IFR. Then for such observers the components of $%
{\bf E}$ and ${\bf B}$, which are not well defined quantities in the ''TT
relativity,'' can be simply replaced by the space components of the
4-vectors $E^{\alpha }$ and $B^{\alpha }$. However, if the LT from $S$ to
another IFR $S^{\prime }$, moving with $V^{\alpha }$ relative to $S$, is
performed, then in $S^{\prime }$ one cannot obtain the usual Maxwell
equations with the 3-vectors ${\bf E}^{\prime }$ and ${\bf B}^{\prime }$
(determined by the AT (\ref{eitran})) from the transformed covariant Maxwell
equations with $E^{\prime \alpha }$ and $B^{\prime \alpha }$.

\section{CONVENTIONAL DEFINITIONS OF THE ELECTROMAGNETIC MOMENTUM AND ENERGY}

In this section we consider the conventional synchronous definitions of the
electromagnetic momentum and energy, (see, e.g., Refs. 1-3, Refs. 15,16).
The total 4-momentum of a charged system can be written as 
\begin{equation}
P_{tot.}^{\mu }=(1/c)\int_{\Sigma }\left[ \Theta ^{\mu \nu }(x)+T^{\mu \nu
}(x)\right] d^{3}\sigma _{\nu }(x),  \label{petot}
\end{equation}
where $d^{3}\sigma ^{\mu }$ are the components of the infinitesimal volume
element of a three-dimensional spacelike hypersurface $\Sigma $, $\Theta
^{\mu \nu }$ is a stress-energy tensor which describes nonelectromagnetic
forces and matter (including the Poincar\'{e} stresses, \cite{poinc} while $%
T^{\mu \nu }$ is the energy-momentum density tensor of the electromagnetic
field. The sum $T_{tot.}^{\mu \nu }=\Theta ^{\mu \nu }(x)+T^{\mu \nu }$ is
divergenceless, $\partial _{\nu }T_{tot.}^{\mu \nu }=0$, causing that the
hyperplane integral (4) is independent of the orientation of that
hyperplane, which enables one to choose $\Sigma $ as the plain $t=a$ in $S$
and $t^{\prime }=b$ in $S^{\prime }$. Both stress-energy tensors in (4),$%
\Theta ^{\mu \nu }$ and $T^{\mu \nu }$, are taken synchronously in $S$ and $%
S^{\prime }$, i.e., at a single time in the observer's frame. But, as shown
in the preceding section, contrary to the assertions in $\left[
2,3,15,16\right] $, (which are based on von Laue's theorem), the integral
over the hyperplane $t=const.$, i.e., $P_{tot.}^{\mu }$ (4), is not a true
4-vector, since the transformation of that integral is - an AT.

\subsection{The Electromagnetic Component}

Using such synchronous definitions, i.e., the choice $t=const.$ for the
hypersurface $\Sigma $ in {\em the observer's frame}, and considering only
the electromagnetic component with $T^{\mu \nu }$ expressed in terms of $%
{\bf E}$ and ${\bf B}$, one finds the conventional form for $P^{\mu }$ of
the field in an IFR $S$%
\begin{equation}
P_{f}^{\mu }=(1/c)\int_{t=a}T^{\mu 0}({\bf r},t)d^{3}x  \label{conpe}
\end{equation}
and the traditional expressions for the field energy and momentum: 
\begin{eqnarray}
cP_{f}^{0} &=&U_{f}=(\varepsilon _{0}/2)\int_{t=a}\left[ {\bf E}^{2}({\bf r}%
,t)+c^{2}{\bf B}^{2}({\bf r},t)\right] d^{3}x=\int_{t=a}u({\bf r},t)d^{3}x, 
\nonumber \\
{\bf P}_{f} &=&\varepsilon _{0}\int_{t=a}{\bf E}({\bf r},t)\times {\bf B}(%
{\bf r},t)d^{3}x=\int_{t=a}{\bf g}({\bf r},t)d^{3}x.  \label{cmpf}
\end{eqnarray}
(Note that $P_{f}^{\mu }$ (5-6) is not a legitimate 4-vector and the
notation with the superscripts is not appropriate, but here we retain such
notation due to historical reasons.) For the determination of $P_{f}^{\prime
}$ and $P_{f}^{\prime 0}$ in $S^{\prime }$ one needs to perform: 1) the
transformations of the integrals, i.e., of the hypersurfaces $t=const.$,
which are the AT, and 2) the AT (1) of ${\bf E}$ and ${\bf B}$. This means
that the conventional synchronous definitions of the electromagnetic energy
and momentum are not in accordance with the ''TT relativity;'' the energy-
momentum (5-6) as a synchronously determined nonlocal physical quantity in
an IFR has nothing to do with the corresponding quantity relevant to
observers in another IFR.

\subsection{The Poincar\'{e} Stresses and the Nonelectromagnetic Component}

In addition, it has to be mentioned that the origin and nature of the
Poincar\'{e} stresses, which are included in $\Theta ^{\mu \nu }$ component,
are unknown, and, in fact, Poincar\'{e} stresses are not measurable physical
quantities. Such a theory with Poincar\'{e}'s stresses is like the theories
which tried in the synchronous formulation of nonlocal physical quantities
to resolve the problem of equilibrium in the special relativity,
particularly the right-angled lever problem, by introducing the fictive
energy current, von Laue's energy current \cite{laue}; (many others repeated
Laue's explanation, see, e.g., Ref. 19). But, as Aranoff $\left[ 20\right] $
stated in his severe criticism of von Laue's explanation: ''The energy
current idea of von Laue has to go the way of phlogiston, and the ether. It
is interesting how man has to invent very fine fluids which carry energy but
which are otherwise unobservable.'' Exactly the same statement is
appropriate for the Poincar\'{e} stresses.

It is interesting to discuss in more detail the explanations given for the
nonelectromagnetic component $\Theta ^{\mu \nu }$ in the theories $\left[
2,3,15,16\right] $. Boyer $\left[ 2\right] $ considers a spherical shell of
charge (as a model of the classical electron) in different IFRs. He
concludes that, in an improper IFR, there is a net transfer of energy and
momentum from the mechanical stabilizing forces to the electromagnetic field
as a consequence of the relativity of simultaneity, i.e., as a result of the
nonsimultaneity of the ''clamping'' of the forces of constraint on the
moving charged shell. On the other hand, according to $\left[ 2\right] $,
there is no such transfer in the rest frame since there the forces of
constraint were applied simultaneously. However, we note that the relativity
of simultaneity is not an intrinsic feature of relativity, but it is a
coordinate dependent effect, i.e., it depends on the kind of synchronization
procedure adopted, and thus on the adopted coordinatization procedure of
Minkowski spacetime. This means that the above mentioned transfer of energy
and momentum $\left[ 2\right] $, will depend on the adopted synchronization
procedure and particularly for the synchronization with absolute
simultaneity, (see, e.g., Ref. 21), the net transfer would need to
disappear. Thus the changes in energy and momentum of the moving charged
shell which own their existence to the relativity of simultaneity are, in
fact, unphysical. Therefore, the assertions given in $\left[ 2\right] $
about the validity and the relativistic correctness of the traditional
definitions (5-6) in every IFR, taken at a single time in that frame, are
unfounded.

Boyer used the relativity of simultaneity to obtain the additional
contributions to the ''pure electromagnetic'' energy and momentum (5-6) for
a moving charged system, i.e., during a transformation from the rest frame
to another IFR. Instead of that the energy increment is explained in $\left[
15\right] $ as arising from the forces of constraint due to work done by
these forces during the Lorentz contraction of a moving charged system. The
additional contribution to the electromagnetic momentum of a moving charged
system is associated in $\left[ 15\right] $ with the energy flow (as von
Laue's energy current) due to forces of constraint. But, as shown in Sec. 2,
the Lorentz contraction is an AT. This fact, together with the use of the
synchronous definitions (5-6), causes that neither the treatment in $\left[
15\right] $ do conform with the ''TT relativity.''

In $\left[ 16\right] $ a parallel plates capacitor and a uniformly charged
spherical shell were considered. In a similar way as in $\left[ 2\right] $
and $\left[ 15\right] $, it is attempted in $\left[ 16\right] $ to obtain $%
T_{tot.}^{\mu \nu }$, which is divergenceless. A special gaseous substance
(a special kind of ''molecules'' that are considered as moving particles or
as continuous gaseous substance) is introduced in order to provide the
additional energy and momentum needed to balance the electromagnetic energy
and momentum. We remark that this gaseous substance is the same kind of
substance as that one involved in von Laue's energy current, i.e., as
Aranoff $\left[ 20\right] $ states: ''....very fine fluids which carry
energy but which are otherwise unobservable.'' Really it is assumed in $%
\left[ 16\right] $ that the mentioned ''molecules'' carry the energy and
momentum but they do not interact with one another and: '' do not change the
dielectric properties of the medium between the plates which is the same as
that of the vacuum.''

We see that in the conventional definitions (4), (5-6) the treatment of the
electromagnetic component includes the AT of the hyperplane $t=const.$ and
the AT of ${\bf E}$ and ${\bf B}$, while the treatment of the
nonelectromagnetic component introduces and uses unphysical quantities.

\section{ROHRLICH'S DEFINITIONS OF THE ELECTROMAGNETIC MOMENTUM 4-VECTOR}

Next, we consider Rohrlich's definition of the electromagnetic momentum
4-vector and the appropriate transformations of the electromagnetic energy
and momentum. Instead of the traditional, synchronous, definitions (5-6) in
the observer's frame Rohrlich $\left[ 22,23,1\right] $ defines the energy
and momentum of the electromagnetic field in a relativistic covariant way.
The general manifestly covariant definition of the electromagnetic
4-momentum in any IFR is given by (4) without the nonelectromagnetic part,
i.e., as 
\begin{equation}
P_{f}^{\mu }=(1/c)\int_{\Sigma }T^{\mu \nu }(x)d^{3}\sigma _{\nu }(x),
\label{pemief}
\end{equation}
with the same meaning of symbols as in (4). The energy momentum tensor $%
T^{\mu \nu }$ is given in terms of $F^{\mu \nu }$ as 
\begin{equation}
T^{\mu \nu }=\varepsilon _{0}\left[ F^{\mu \alpha }F_{\alpha }^{\nu
}+(1/4)g^{\mu \nu }F_{\alpha \beta }F^{\alpha \beta }.\right]   \label{temif}
\end{equation}
Rohrlich $\left[ 1,22,23\right] $ defined the electromagnetic
energy-momentum for a system in uniform motion as the Lorentz boosted rest
frame, and the hyperplane of integration is specified to be the plane in
which the system is at rest. In such a way it is achieved that $P_{f}^{\mu }$
is a 4-vector even though there are sources present. In the IFR $S_{(0)}$ in
which the system is at rest Rohrlich chooses the hypersurface $\Sigma $ in
(7) to be the plane $\Sigma _{(0)}=t_{(0)}=const.$. {\em The same }$\Sigma $%
{\em \ is considered from all other IFRs as required by the covariant
approach,} i.e., in some IFR $S$ obtained by the Lorentz transformation $L$
from $S_{(0)}$, $\Sigma $ in (7) is $L\Sigma _{(0)}$. $d^{3}\sigma _{\nu }$
in an IFR $S$ can be written as $d^{3}\sigma ^{\nu }=n^{\nu }d^{3}\sigma $,
where $n^{\nu }$ being a timelike vector normal to the hyperplane $\Sigma $, 
$n^{\nu }=\Gamma (1,{\bf \beta })$, (${\bf \beta =V}/c$, ${\bf V}$ is the
3-velocity of the frame $S$ relative to $S_{(0)}$). In $S_{(0)}\quad
n_{(0)}^{\mu }=(1,{\bf 0})$. $d^{3}\sigma $ is an invariant and it is $%
=dV_{(0)}$, where $dV_{(0)}$ is the infinitesimal element of the 3D spatial
volume $V_{(0)}$, i.e., of the 3D hyperplane $t_{(0)}=const.$, in the rest
frame $S_{(0)}$ of the system. {\em We see that in Rohrlich's approach one
always integrates over the hyperplane which is the transformed three space
of the rest frame.} Obviously, according to the construction, i.e., since
always the same physical quantity is considered from different IFRs and only
the TT are used, it is found that the quantity $P_{f}^{\mu }$ (7) is a
legitimate 4-vector. However, when Rohrlich explicitly calculates (7) for
specific physical system he also writes, as all others, the electromagnetic
energy and momentum densities in terms of the 3-vectors ${\bf E}$ and ${\bf B%
}$ and uses the AT (1).

\section{ELECTROMAGNETIC MOMENTUM 4-VECTOR IN \\TERMS OF $E^{\alpha }$ AND $%
B^{\alpha }$}

In order to remove the last element from the theory that is not in
accordance with the ''TT relativity,'' we write $T^{\mu \nu }$ and $%
P_{f}^{\mu }$ in terms of covariant quantities $E^{\alpha }$ and $B^{\alpha
} $. Then we compare so obtained $T^{\mu \nu }$ and $P_{f}^{\mu }$ with
traditional definitions of the electromagnetic energy and momentum (5-6) and
with Rohrlich's expressions (3.23) and (3.24) in $\left[ 23\right] $ or with
the expressions for $P_{e}^{0}$ and $P_{e}^{k}$ obtained in $\left[ 1\right] 
$ for the choice (II).

Using the relations (3) we express $F^{\alpha \beta }$ and thus also $T^{\mu
\nu }$ (7) in terms of $E^{\alpha }$ and $B^{\alpha }$. The obtained
covariant expression for the symmetric energy-momentum density tensor $%
T^{\mu \nu }$ is the following 
\begin{eqnarray}
T^{\mu \nu } &=&\varepsilon _{0}\left[ -((g^{\mu \nu }/2)+v^{\mu }v^{\nu
}/c^{2})(E_{\alpha }E^{\alpha }+c^{2}B_{\alpha }B^{\alpha })\right. 
\nonumber \\
&&+E^{\mu }E^{\nu }+c^{2}B^{\mu }B^{\nu }  \label{teco} \\
&&+(1/c)\varepsilon ^{\mu \alpha \beta \gamma }B_{\alpha }(v^{\nu }v_{\gamma
}E_{\beta }-E^{\nu }v_{\beta }v_{\gamma })  \nonumber \\
&&\left. +(1/c)\varepsilon ^{\nu \alpha \beta \gamma }B_{\alpha }(v^{\mu
}v_{\gamma }E_{\beta }-E^{\mu }v_{\beta }v_{\gamma })\right]  \nonumber
\end{eqnarray}
Introducing (9) into (7) one finds the explicit expression for the
covariantly defined electromagnetic 4-momentum $P_{f}^{\mu }$. All parts of
the expression (7) with $T^{\mu \nu }$ from (9) are covariantly defined
quantities which transform according to the TT, i.e., according to the LT;
in another IFR $S^{\prime }$ moving with the 4-velocity $V^{\mu }$ relative
to $S$ the 4-momentum (7) with $T^{\mu \nu }$ defined by (9) will have the
same form but with primed quantities replacing the unprimed ones.

\subsection{$P_{f}^{\mu }$ in $S_{(0)}$ and $S$}

Let us now use the above manifestly covariant expression for $P_{f}^{\mu }$
with $E^{\alpha }$ and $B^{\alpha }$ to examine some specific cases
considered in $\left[ 1\right] $ and $\left[ 2\right] $. First we write $%
T_{(0)}^{\mu \nu }$ and $P_{f(0)}^{\mu }$ in $S_{(0)}$, the rest frame of
the charged sphere, i.e., when $u^{\alpha }$ (the 4-velocity of the charge)
is $u^{\alpha }=(c,{\bf 0)}$, whence $n_{(0)\nu }=(-1,{\bf 0)}$. The
observers who measure $E_{(0)}^{\alpha }$ and $B_{(0)}^{\alpha }$ are taken
to be at rest in $S_{(0)}$, and thus $v^{\alpha }$ (the 4-velocity of the
observers) is $v^{\alpha }=u^{\alpha }=(c,{\bf 0)}$. For such observers in $%
S_{(0)}$ one finds $E_{(0)}^{0}=0$, $B_{(0)}^{0}=0$ and only $%
E_{(0)}^{i}\neq 0$. From 
\begin{equation}
P_{f(0)}^{\mu }=(1/c)\int_{\Sigma _{(0)}}T_{(0)}^{\mu \nu }n_{(0)\nu
}dV_{(0)},  \label{penul}
\end{equation}
we find that 
\begin{equation}
P_{f(0)}^{0}=-(1/c)\int_{\Sigma _{(0)}}T_{(0)}^{00}dV_{(0)},\quad
P_{f(0)}^{i}=-(1/c)\int_{\Sigma _{(0)}}T_{(0)}^{i0}dV_{(0)}=0,  \label{des}
\end{equation}
where from (9) the components of $T_{(0)}^{\mu \nu }$ can be expressed in
terms of $E_{(0)}^{\alpha }$ and $B_{(0)}^{\alpha }$ as $%
T_{(0)}^{00}=-u_{E(0)}=-(\varepsilon _{0}/2)(E_{(0)}^{i}E_{(0)i})$, $%
T_{(0)}^{i0}=T_{(0)}^{0i}=0$, $T_{(0)}^{11}=\varepsilon
_{0}(E_{(0)}^{1})^{2}-u_{E(0)}$, and similarly for $T_{(0)}^{22}$ and $%
T_{(0)}^{33}$ with indexes 2 and 3 replacing the index 1, and $%
T_{(0)}^{ij}=T_{(0)}^{ji}=\varepsilon _{0}E_{(0)}^{i}E_{(0)}^{j}$, with $%
i\neq j$. $E_{(0)}^{i}$ are the space components of the 4-vector $%
E_{(0)}^{\alpha }$ and they correspond to the usual Coulomb field. Thus the
self-energy due to the Coulomb field $U_{f(0)}$ and the space part $%
P_{f(0)}^{i}$ are 
\begin{eqnarray}
U_{f(0)} &=&cP_{f(0)}^{0}=\int_{\Sigma _{(0)}}u_{E(0)}dV_{(0)}=(\varepsilon
_{0}/2)\int_{\Sigma _{(0)}}E_{(0)}^{i}E_{(0)i}dV_{(0)},  \nonumber \\
P_{f(0)}^{i} &=&0  \label{upeo}
\end{eqnarray}
In order to find the 4-momentum $P_{f}^{\mu }$, Eq. (7), in another IFR $S$
moving with the 4-velocity $V^{\alpha }=(\Gamma c,\Gamma V,0,0)$ relative to 
$S_{(0)}$ one can either transform $P_{f(0)}^{\mu }$ as a 4-vector from $%
S_{(0)}$ to $S$, or to transform all quantities on the right-hand side of
(10) from $S_{(0)}$ to $S$. The same result is obtained and it is 
\begin{equation}
P_{f}^{0}=\Gamma P_{f(0)}^{0},\quad P_{f}^{1}=-\beta \Gamma
P_{f(0)}^{0},\quad P_{f}^{2}=P_{f}^{3}=0.  \label{pfj}
\end{equation}
Note that the same family of observers who measures $E_{(0)}^{\alpha }$ and $%
B_{(0)}^{\alpha }$ in $S_{(0)}$ is considered in all other IFRs. Wee see
that, when referred to the invariant 3D integration volume $dV_{(0)}$, the
energy density $u_{E}$ in the IFR $S$ moving with $V^{\alpha }=(\Gamma
c,\Gamma V,0,0)$ relative to $S_{(0)}$ is $\Gamma u_{E(0)}$, contrary to the
results in $\left[ 1,22,23\right] $.

\subsection{The (''4/3'') Factor}

We can use these results to discuss the famous ''4/3'' factor appearing in
the problem of the electromagnetic mass of the classical electron, (see,
e.g., Refs. 1-3). Let us suppose that in the rest system $S_{(0)}$ the whole
mass $m$ of the electron (considered as a sphere of radius $R$ with a
uniform surface charge density) is due to electrostatic energy of the field.
Using the traditional, synchronous, definitions (5-6), one finds that in $%
S_{(0)}$ $cP_{f(0)}^{0}=mc^{2}=U_{f(0)}$, and ${\bf P}_{f(0)}=0$. In an IFR $%
S$ in which the particle moves with the velocity ${\bf u}$ one obtains from
(5-6) $P_{f}^{0}=\Gamma m(1+{\bf u}^{2}/3)$, and ${\bf P}_{f}=(4/3)\Gamma m%
{\bf u}$, (see eqations (11) and (12) in Rohrlich's criticism $\left[
1\right] $ of the work $\left[ 2\right] $). We see that the spurious 4/3
factor appears in ${\bf P}_{f}$. Of course, as already said, the quantities $%
P_{f}^{0}$ and ${\bf P}_{f}$ do not form a 4-vector. Because of that $%
f_{coh}^{\mu }$, the force density that provides the Poincar\'{e} stresses,
is introduced into the theory. Then $P_{coh}^{0}$ and ${\bf P}_{coh}$ are
calculated by means of $f_{coh}^{\mu }$ in such a way to give that the sum
of two false 4-vectors $P_{f}^{\mu }$ and $P_{coh}^{\mu }$ is a hyperplane
integral independent of the orientation of that hyperplane. Applying von
Laue's theorem it is concluded by the proponents of the synchronous
definitions that such sum is a legitimate 4-vector, (see, e.g., Ref. 1 Eqs.
(14) and (15)). However, despite the fact that the sum $P_{f}^{\mu
}+P_{coh}^{\mu }$ can be written in the form of a 4-vector, i.e., as $%
mv^{\mu }$, (see equation (15) in $\left[ 1\right] $), this quantity is not
a true 4-vector. As we have already shown this sum does not refer to the
same quantity considered in different IFRs; the plains $t_{(0)}=a$ in $%
S_{(0)}$ and $t=b$ in $S$ are not related by the LT than by the AT, and the
relations (1) connecting ${\bf E}$ and ${\bf B}$ in $S$ and ${\bf E}_{(0)}$
and ${\bf B}_{(0)}$ in $S_{(0)}$ are not the LT of the same quantities from $%
S_{(0)}$ to $S$, than they are also the AT.

In contrast to the synchronous definitions considered in $\left[
2,3,15,16\right] $, Rohrlich's expressions for $P_{f}^{0}$ and $P_{f}^{k}$
in $S$, derived in the Appendix in $\left[ 1\right] $, give that $P_{f}^{\mu
}$ alone can be written as $mv^{\mu }$, and accordingly it is also concluded
in $\left[ 1\right] $, by the use of von Laue's theorem, that $P_{f}^{\mu }$
is a true 4-vector. In Rohrlich's approach there is no spurious 4/3 factor
in $P_{f}^{\mu }$, and, in contrast to Boyer's approach, Rohrlich's $\Sigma $
in $S$ is correctly determined as $L\Sigma _{(0)}$. But, as already said,
neither Rohrlich's $P_{f}^{\mu }$ is a legitimate 4-vector, since he uses
the AT of ${\bf E}$ and ${\bf B}$. The use of ${\bf E}$ and ${\bf B}$ and
the AT (1) in $\left[ 1\right] $ instead of the 4-vectors $E^{\alpha }$ and $%
B^{\alpha }$ and their LT causes the difference between the results obtained
in $\left[ 1\right] $ for the energy and momentum densities and for $%
P_{f}^{0}$ and $P_{f}^{k}$, and the corresponding expressions obtained here,
Eqs. (12) and (13).

>From (11,12) and (13) we see that in the covariant approach presented here
the spurious factor 4/3 does not appear and that ''pure electromagnetic'' $%
P_{f}^{\mu }$ is a legitimate 4-vector; it refers to the same quantity in
all IFRs since all parts of it are Lorentz transformed when going from IFR $%
S_{(0)}$ to some IFR $S$.

\subsection{Rohrlich's $P_{f}^{0}$ and $P_{f}^{k}$ from $P_{f}^{\mu }$ with $%
E^{\alpha }$ and $B^{\alpha }$}

The expressions for $P_{f}^{\mu }$ (written with $E^{\alpha }$ and $%
B^{\alpha }$) corresponding to Rohrlich's relations for $P_{f}^{0}$ and $%
P_{f}^{k}$ with ${\bf E}$ and ${\bf B}$ $\left[ 1\right] $, can be obtained
in the following way. Let the IFR $S$ be the frame in which the particle
moves with the 4-velocity $u^{\alpha }=(\gamma _{u}c,\gamma _{u}u,0,0)$, and
therefore the unit 4-vector $n^{\mu }$ is $n^{\mu }=(\gamma _{u},\gamma
_{u}\beta _{u},0,0)$. The 4-momentum is given by (7). Using these relations
we write the components of $P_{f}^{\mu }$ as 
\begin{eqnarray}
P_{f}^{0} &=&-(\gamma _{u}/c)\int dV_{(0)}\left[ T^{00}-\beta
_{u}T^{10}\right] ,  \nonumber \\
P_{f}^{i} &=&-(\gamma _{u}/c)\int dV_{(0)}\left[ T^{0i}-\beta
_{u}T^{1i}\right] .  \label{mome}
\end{eqnarray}
Further, let the observers who measure the fields $E^{\alpha }$ and $%
B^{\alpha }$ in $S$ are at rest in $S$, i.e., their 4-velocity $v^{\alpha }$
is $v^{\alpha }=(c,{\bf 0)}$. For such observers $E^{0}=B^{0}=0$. Then from
(9) we find that $T^{\mu \nu }$ can be written in terms of the components of 
$E^{\alpha }$ and $B^{\alpha }$ as 
\begin{eqnarray}
T^{00} &=&-u_{E}=-(\varepsilon _{0}/2)(E^{i}E_{i}+c^{2}B^{i}B_{i}), 
\nonumber \\
T^{0i} &=&-\varepsilon _{0}c\varepsilon _{ijk}E^{j}B^{k},  \label{tete} \\
T^{11} &=&\varepsilon _{0}((E^{1})^{2}+c^{2}(B^{1})^{2})-u_{E},  \nonumber \\
T^{1n} &=&\varepsilon _{0}(E^{1}E^{n}+c^{2}B^{1}B^{n}),\quad n=2,3. 
\nonumber
\end{eqnarray}
When $T^{\mu \nu }$ from (15) is introduced into (14) then $P_{f}^{0}$ and $%
P_{f}^{i}$ seem like $P_{e}^{0}$ and $P_{e}^{i}$ for the choice (II) in.$%
\left[ 1\right] .$ {\em But the 4-vectors }$E^{\alpha }${\em \ and }$%
B^{\alpha }${\em \ in (15) are measured by the observers at rest in S},
i.e., whose velocity is $v^{\alpha }=(c,{\bf 0)}$, which means that these $%
E^{\alpha }$ and $B^{\alpha }$ are not the LT of the previously mentioned $%
E_{(0)}^{\alpha }$ and $B_{(0)}^{\alpha }$ (for which $v^{\alpha }=(c,{\bf 0)%
}$ in $S_{(0)}$). Thus we find that, contrary to the derivation in the
Appendix of $\left[ 1\right] $, Eq. (14) with $T^{\mu \nu }$ determined by
(15) is not the LT of $P_{f(0)}^{\mu }$ (11,12). The LT of $P_{f(0)}^{\mu }$
are actually given by (13), as it is shown above. If one performs the same
procedure as in the Appendix of $\left[ 1\right] $ expressing the 4-vectors $%
E^{\alpha }$ and $B^{\alpha }$ in $S$ by means of $E_{(0)}^{\prime \alpha }$
and $B_{(0)}^{\prime \alpha }$, the 4-vectors in the rest frame $S_{(0)}$ of
the charged sphere, which are connected by the LT with $E^{\alpha }$ and $%
B^{\alpha }$, then one does not find $P_{e}^{0}$ and $P_{e}^{i}$ obtained
for the choice (II) in.$\left[ 1\right] $. The 4-vectors $E_{(0)}^{\prime
\alpha }$ and $B_{(0)}^{\prime \alpha }$ are not equivalent to the
previously considered (in connection with (11-13)) 4-vectors $%
E_{(0)}^{\alpha }$ and $B_{(0)}^{\alpha }$; the former refer to the electric
and magnetic fields which are measured in $S$ by the observers at rest in $S$
and then Lorentz transformed to $S_{(0)}$, while the latter refer to the
electric and magnetic fields which are measured directly in $S_{(0)}$ by the
observers at rest in $S_{(0)}$. The preceding discussion reveals that the
difference between the results in $\left[ 1\right] $ and in this paper is,
as already said, a consequence of the use of the 3-vectors ${\bf E}$ and $%
{\bf B}$ and the AT (1) in $\left[ 1\right] $, and the use of the 4-vectors $%
E^{\alpha }$ and $B^{\alpha }$ and their LT in this paper.

\section{SOME RECENT TREATMENTS OF THE ELECTROMAGNETIC MOMENTUM AND ENERGY}

In this section we consider some recent treatments of the electromagnetic
energy and momentum.

\subsection{Romer's Question and Answers}

Recently Romer $\left[ 24\right] $ revived the question of the correct
expressions for the electromagnetic field momentum and energy in the case of
''bound'' fields, fields that are tied to their sources. He, and many others
(see references in $\left[ 24\right] $), uses the traditional synchronous
definitions (5-6). As we have already shown the relations (5-6) contain the
AT of the hyperplane $t=const.$ and the AT of ${\bf E}$ and ${\bf B}$.
Different answers to this question have been given in $\left[ 25\right] $.
Neither the answers $\left[ 25\right] $ to the question in $\left[ 24\right] 
$ are in a complete agreement with the ''TT relativity.'' First, they also
work with ${\bf E}$ and ${\bf B}$ and their AT (1), and base their
conclusions on von Laue's theorem.

\subsection{Schwinger's Consideration of the ''4/3'' Problem}

We have to mention an interesting consideration of the electromagnetic
energy and momentum and the electromagnetic mass given in $\left[ 26\right] $
by Schwinger. He also uses the synchronous definition of $P_{f}^{\mu }$ (5).
In difference to the works $\left[ 2,3,15,16\right] $ he does not deal with
Poincar\'{e}'s stresses, but changes the definition of the electromagnetic
energy-momentum tensor $T^{\mu \nu }$ (8). As it is already said $T^{\mu \nu
}$ (8) is not divergence-free and therefore the hyperplane integral of $%
T^{\mu \nu }$ (7), is dependent of the orientation of the hyperplane.
Schwinger construes, but only for a class of fields and currents, associated
with uniform motion, a new, conserved, divergenceless, energy-momentum
tensor $T_{Sch.}^{\mu \nu }$ of the electromagnetic field. Since $\partial
_{\nu }T_{Sch.}^{\mu \nu }=0$ it is achieved in $\left[ 26\right] $ that the
hyperplane integral (7), (with $T_{Sch.}^{\mu \nu }$ replacing $T^{\mu \nu }$
(8)), is independent of the orientation of that hyperplane, and this is
considered by Schwinger too as a necessary and sufficient condition that the
integral (7) is a true 4-vector, i.e., he also accepts von Laue's theorem as
that it is a correct one from the ''TT viewpoint.'' Then choosing that $%
\Sigma $ in (7) is the plane $t=a$ in some IFR $S$ the integral (7) becomes
the synchronous definition of $P_{f}^{\mu }$ (5). Also it has to be noted
that the tensor $T_{Sch.}^{\mu \nu }$ is not unique, and for the same
field-current distribution one can have different $T_{Sch.}^{\mu \nu }$,
e.g., the tensor (1) equation (42) and the tensor (2) equation (44) in $%
\left[ 26\right] $. Using different $T_{Sch.}^{\mu \nu }$ ((1) and (2) in $%
\left[ 26\right] $) two different ''covariant'' versions of the concept of
electromagnetic mass were obtained in $\left[ 26\right] $. We note that the
same remarks as for the usual synchronous definitions hold also here; $%
t_{(0)}=const.$ in $S_{(0)}$ and $t=a$ in $S$ are not related by the LT than
by the AT. Further, the energy and momentum in $\left[ 26\right] $ are
ultimately expressed in terms of ${\bf E}$ and ${\bf B}$, (see equations
(43), (45) and (46) in $\left[ 26\right] $). Thus, contrary to the
assertions in $\left[ 26\right] $, we conclude that even though the energy
and momentum defined by (5) (with $T_{Sch.}^{\mu 0}$ instead of $T^{\mu 0}$)
transform like a 4-vector (see equations (62-63) in $\left[ 26\right] $),
these equations, which are derived from $P_{f}^{\mu }$ (5), do not define a
true 4-vector, i.e., they do not covariantly define the energy and momentum
of the ''bound'' electromagnetic field. The AT are used in the derivation in 
$\left[ 26\right] $, whence one concludes that neither the equations (62)
based on rest mass $m^{(2)}$ nor the relations (63) based on rest mass $%
m^{(1)}$ do refer to the same physical quantity considered from different
IFRs.

\subsection{Some Other Treatments}

Recently the electromagnetic mass derived from the self-force and the 4/3
factor were discussed in $\left[ 27\right] $. These works will not be
considered here, but we only mention that, ultimately, they use the
synchronous definitions and the 3-vectors ${\bf E}$ and ${\bf B}$.

In the recent work $\left[ 14\right] $ the covariant Majorana formulation of
electrodynamics is constructed. There, the covariant expression (with the
4-vectors $E^{\alpha }$ and $B^{\alpha }$) for $T^{\mu \nu }$ is obtained,
and it is equal to our Eq. (9). But, as we have said, Esposito $\left[
14\right] $ considers the fields $E^{\alpha }$ and $B^{\alpha }$ as
''auxiliary'' fields, while the 3-vector fields ${\bf E}$ and ${\bf B}$ are
considered as physical fields. The situation is just the opposite in our
alternative covariant approach.

The preceding discussion indicates that the correct definitions from the
''TT viewpoint'' of the electromagnetic 4-momentum have to contain only
covariantly defined quantities, including the 4-vectors $E^{\alpha }$ and $%
B^{\alpha }$ instead of the usual 3-vectors ${\bf E}$ and ${\bf B}$, and the
TT of 4D tensor quantities instead of the AT. Such relations are the
expressions (7-15) in this paper.

\section{DISCUSSION AND\ CONCLUSIONS}

The fundamental difference between the apparent and true transformations of
physical quantities, which is previously mainly overlooked, enabled us to
reveal that the usual formulation of electrodynamics with the 3-vectors $%
{\bf E}$ and ${\bf B}$ is not in agreement with the ''TT relativity.''
Different definitions of the electromagnetic energy and momentum are shown
to be invalid from the ''TT viewpoint'' since they contain either the AT of
volume, or the AT of ${\bf E}$ and ${\bf B}$, or both of them. We have
constructed a covariant formulation of electrodynamics with the 4-vectors $%
E^{\alpha }$ and $B^{\alpha }$ equally as valid as the usual covariant
approach with $F^{\alpha \beta }$. The covariant expression for the
symmetric energy-momentum density tensor $T^{\mu \nu }$ is obtained by means
of $E^{\alpha }$ and $B^{\alpha }$. The electromagnetic 4-momentum $%
P_{f}^{\mu }$ with $E^{\alpha }$ and $B^{\alpha }$ is shown to be a true
4-vector. $P_{f}^{\mu }$ refers to the same quantity considered in different
IFRs, since all parts of it are transformed by the LT from an IFR $S$ to
relatively moving IFR $S^{\prime }$.

We emphasize that this approach with $E^{\alpha }$ and $B^{\alpha }$ is not
restricted to the classical electrodynamics but refers in the same measure
to the quantum electrodynamics. The 3-vectors ${\bf E}$ and ${\bf B}$,
(whose transformations are the AT), can be replaced by the 4-vectors $%
E^{\alpha }$ and $B^{\alpha }$, (which transform according to the TT), in
the quantum electrodynamics as well. The use of $E^{\alpha }$ and $B^{\alpha
}$ in the quantization of the electromagnetic field does have important
advantages: 1) $E^{\alpha }$ and $B^{\alpha }$ are covariant quantities, 2)
one does not need to use the intermediate electromagnetic 4-potential $%
A^{\mu }$, and thus dispenses with the need for gauge conditions.\bigskip
\medskip 

\noindent {\bf ACKNOWLEDGMENTS\medskip }

I am indebted to Prof. F. Rohrlich for suggestions how to improve the
presentation of the paper and, for the same reason, to an anonymous referee.


\begin{thebibliography}{99}
\bibitem{RochD}  F. Rohrlich, {\it Phys. Rev. D} {\bf 25}, 3251 (1982).

\bibitem{Boyer}  T.H. Boyer, {\it Phys. Rev. D} {\bf 25}, 3246 (1982).

\bibitem{camp}  I. Campos and J.L. Jim\'{e}nez, {\it Phys. Rev. D} {\bf 33},
607 (1986).

\bibitem{rochnc}  F. Rohrlich, {\it Nuovo Cimento B} {\bf 45}, 76 (1966).

\bibitem{gamba}  A. Gamba, {\it Am. J. Phys.} {\bf 35}, 83 (1967).

\bibitem{ivez12}  T. Ivezi\'{c}, preprint SCAN-9802018; preprint
SCAN-9810078 (on the CERN server).

\bibitem{gron}  \O . Gr\o n, {\it Am. J. Phys.} {\bf 46}, 249 (1978); T.
Ivezi\'{c}, {\it Phys. Rev. E} {\bf 52}, 5505 (1995).

\bibitem{einst}  A. Einstein, {\it Ann. Physik} {\bf 17}, 891 (1905), tr. by
W. Perrett and G.B. Jeffery, in {\it The principle of relativity }(Dover,
New York).

\bibitem{misner}  C.W. Misner, K.S. Thorne and J.A. Wheeler, {\it Gravitation%
}, (Freeman, San Francisco, 1970).

\bibitem{pauli}  W. Pauli, {\it The theory of relativity} (Pergamon Press,
London, 1958); S.A. Teukolsky, {\it Am. J. Phys.} {\bf 64}, 1104 (1996).

\bibitem{jacks}  J. D. Jackson, {\it Classical Electrodynamics}, 2nd edn.
(Wiley, New York, 1977).

\bibitem{wald}  R. M. Wald, {\it General relativity} (The University of
Chicago Press, Chicago, 1984); D.A.T. Vanzella, G.E.A. Matsas, H.W. Crater, 
{\it Am. J. Phys.} {\bf 64}, 1075 (1996).

\bibitem{anders}  J.L. Anderson and J.W. Ryon, {\it Phys. Rev.} {\bf 181},
1765 (1969); P. Hillion, {\it Phys. Rev. E} {\bf 48}, 3060 (1993).

\bibitem{espos}  S. Esposito, {\it Found. Phys.} {\bf 28}, 231 (1998).

\bibitem{singal}  A.K. Singal, {\it J. Phys. A} {\bf 25}, 1605 (1992).

\bibitem{comay}  E. Comay, {\it Z. Naturforsch.} {\bf 46a}, 377 (1991).

\bibitem{poinc}  H. Poincar\'{e}, {\it Rend. Circ. Mat. Palermo} {\bf 21},
129 (1906).

\bibitem{laue}  M. von Laue, {\it Phys. Zeits.} {\bf 12}, 1008 (1911).

\bibitem{panof}  W. Pauli, The theory of relativity (Pergamon Press, London,
1958); W.K.H. Panofsky and M. Phillips, {\it Classical electricity and
magnetism}, 2nd edn. (Addison-Wesley, Reading, PA, 1962).

\bibitem{aranof}  S. Aranoff, {\it Nuovo Cimento B} {\bf 10}, 155 (1972).

\bibitem{caval}  G. Cavalleri and C. Bernasconi, {\it Nuovo Cimento B} {\bf %
104}, 545 (1989); C. Leubner, K. Aufinger and P. Krumm, {\it Eur. J. Phys.} 
{\bf 13}, 170 (1992).

\bibitem{rohclas}  F. Rohrlich, {\it Classical charged particles},
(Addison-Wesley, Reading, MA, 1965).

\bibitem{rocam}  F. Rohrlich, {\it Am. J. Phys.} {\bf 38}, 1310 (1970).

\bibitem{romer}  R.H. Romer, {\it Am. J. Phys.} {\bf 63}, 777 (1995).

\bibitem{donald}  K. McDonald, {\it Am. J. Phys.} {\bf 64}, 15 (1996); F.
Rohrlich, {\it Am. J. Phys.} {\bf 64}, 16 (1996); B.R. Holstein, {\it Am. J.
Phys.} {\bf 64}, 17 (1996).

\bibitem{schwin}  J. Schwinger, {\it Found. Phys.} {\bf 13}, 373 (1983).

\bibitem{frenk}  J. Frenkel, {\it Phys. Rev. E} {\bf 54}, 5859 (1996); H.
Kolbenstvedt, {\it Phys. Lett. A} {\bf 234}, 319 (1997).
\end{thebibliography}
\end{document}